\documentclass[twocolumn,showpacs,preprintnumbers,amsmath,amssymb]{revtex4}

\usepackage{graphicx}
\usepackage{dcolumn}
\usepackage{bm}

\begin{document}

\title{Anisotropic magnetization and sign change of dynamic susceptibility in Na$_{0.85}$CoO$_2$ single crystal}

\author{Jong-Soo Rhyee$^{1,3,*}$, J. B. Peng$^{1,2}$, C. T. Lin$^{1,\dag}$, and S. M. Lee$^{3}$}
\affiliation{$^{1}$Max-Plank-Institute for Solid State Research,
Heisenbergstrasse 1,
Stuttgart, Germany D-70569. \\
$^{2}$Kunming University of Science and Technology, Yunnan 650093,
PR China. \\
$^3$Advanced Materials Lab., Samsung Advanced Institute of
Technology, Suwon 440-600, Korea.}


\begin{abstract}
The DC and AC magnetic susceptibilities of Na$_{0.85}$CoO$_2$ single
crystals were measured for the different crystal orientations of
$H\parallel$(ab)- and $H\parallel$(c)-axis. The DC-magnetic
susceptibility for $H\parallel$(c)-direction exhibited the
antiferromagnetic transition at $T_N=$ 22 K. The thermal hysteresis
between the zero-field-cooled (ZFC) and the field-cooled (FC)
magnetization below $T_N$ and the large frustration parameter
indicated the spin frustration along the $c$-axis. For an applied
magnetic field in $H\parallel$(ab)-plane, the DC magnetic
susceptibility exhibited the logarithmic divergent behavior at low
temperatures ($T\leq$ 6.8 K). This could be understood by the
impurity spin effect, dressed by the spin fluctuation. From the AC
magnetic susceptibility measurements, the real part of the
AC-susceptibility for $H\parallel$(ab) exhibited the spin glass-like
behavior at low temperatures ($T\leq $ 4 K). Remarkably, for an
applied AC magnetic field with $H\parallel$(c)-axis, the sign of the
AC magnetic susceptibility changed from a positive to a negative
value with increasing AC magnetic field frequency ($f\geq$ 3 kHz) at
low temperatures ($T\leq$ 7 K). We interpret the sign change of AC
magnetic susceptibility along the {\it c}-axis in terms of the
sudden sign reversal of the phase difference $\phi$ from in-phase to
out-of-phase response with an applied AC magnetic field in the
AC-susceptibility phase space. \\

\noindent $^{\ast}$Corresponding author: js.rhyee@samsung.com (J.S.R.) \\
$^{\dag}$Request for materials should be addressed to:
ct.lin@fkf.mpg.de (C.T.L.)

\end{abstract}

\maketitle
\newpage

\section{Introduction}
Since the discovery of superconductivity ($T_c\approx$ 4.5 K) in the
hydrated compound of Na$_x$CoO$_2 y$H$_2$O ($x\approx$ 0.35,
$y\approx$ 1.3), much attention has been devoted to the Na cobaltate
series compounds of Na$_x$CoO$_2$.\cite{Takada03} Nonhydrated
compounds of the Na cobaltate system Na$_x$CoO$_2$ (0.3 $\leq x
\leq$ 1.0) exhibit various physical properties depending upon the Na
nonstoichiometry, such as high thermoelectric
power,\cite{Wang03,Lee06} quantum criticality,\cite{Rivadulla06}
charge ordering,\cite{Foo04,Mukhamedshin04} spin density
wave,\cite{Bayrakci05} magnetic polaron,\cite{Bernhard04} {\it etc}.
The variety of physical properties exhibited by this system may
originate from the strong correlation among electron, spin, and
orbital degrees of freedom. Since few years, intense investigations
have been carried out on the magnetic ground state of
$\gamma-$Na$_x$Co$_y$O$_2$ (0.75 $\leq x/y \leq$ 0.85, P2 phase)
compounds. In the neutron scattering experiments of Na$_{x}$CoO$_2$
($x=$ 0.75 and 0.85) compounds, ferromagnetic interaction was
revealed within the CoO$_2$ layers while antiferromagnetic (AF)
ordering was observed perpendicular to the CoO$_2$ layers (A-type
AF).\cite{Helme05,Helme06} Bulk antiferromagnetism was observed at
the N\'{e}el temperature $T_N=$ 19.8 K along the $c$-axis in a
Na$_{0.82}$CoO$_2$ crystal with two-dimensional antiferromagnetic
fluctuation near 30 K. In addition, from the susceptibility,
specific heat, and muon spin rotation measurements, a spin density
wave was observed, which is an indication of the two-dimensional
spin state.\cite{Bayrakci04,Bayrakci05}

The two-dimensional characteristics of magnetic interaction and
various magnetic and electronic ground states are mainly due to the
layered structure of the CoO$_2$ layer and the unconventional spin
state of the Co ion in Na$_{x}$CoO$_2$. The crystal structure of the
CoO$_2$ sublattice is a triangular network of edge-sharing oxygen
octahedra.\cite{Jorgensen03} The magnetic ground state of
Na$_x$CoO$_2$ is very sensitive with respect to the Na
nonstoichiometry. It may be affected by both the charge or the spin
state of the Co ion and the crystal structure of the CoO$_2$
sublattices. The detailed crystal structure variation of
Na$_x$CoO$_2$ is very complex with respect to Na concentration
$x$.\cite{Viciu06} In particular, the Na$_{0.85}$CoO$_2$ crystal is
in the upper bound region of the $\gamma-$phase $H2-$type structure.
It had the highest value of the Seebeck coefficient in addition to
exhibiting the strong spin frustration behavior
expected.\cite{Lee06} Significant dynamical spin behavior is usually
observed in a strong spin frustration system. Because the strong
spin frustration was expected, the measurement of the dynamical spin
property would be helpful in understanding the frustration behavior
of the Na$_{0.85}$CoO$_2$ compound.

Therefore, we measured the anisotropic DC and AC magnetic
susceptibilities of a Na$_{0.85}$CoO$_2$ single crystal in order to
study the anisotropic behaviors of the static and dynamic spin
response of the Co ion. From the measurement of the DC magnetic
susceptibility, it was found that there existed a thermal hysteresis
between the zero-field-cooled (ZFC) and the field-cooled (FC)
magnetization at low temperatures ($T\leq$ 20 K), which implied an
unstable magnetic background. On the other hand, an
antiferromagnetic (AF) transition was observed at $T_N\approx$ 22 K
for $(M/H)_c$ along the $H\parallel$(c) direction. Exotic behaviors
of $(M/H)_{ab}$ - logarithmic divergent DC magnetic susceptibility
at low temperatures ($T\leq$ 7 K) and power law dependence
$(M/H)_{ab}\propto T^{\alpha}$ ($\alpha=-$0.078) at high
temperatures ($T\geq$ 100 K) - were also observed. In addition, from
the measurement of the AC magnetic susceptibility, we observed an
unconventional sign change from positive to negative of the dynamic
spin susceptibility $\chi_{c}(f,T)\equiv dM_c/dH$ with respect to
frequencies and temperatures. In this paper, we have discussed the
unusual anisotropic magnetic state and AC frequency response of the
dynamic magnetic susceptibility of the Na$_{0.85}$CoO$_2$ crystal.

\section{Experimental Details}
\begin{figure}
\includegraphics [width=0.5\textwidth]{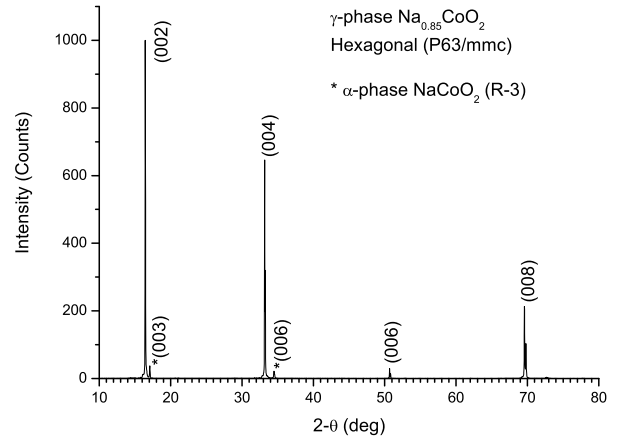}
\caption{X-ray diffraction pattern of an Na$_{0.85}$CoO$_2$ single
crystal. Major sharp peaks (00{\it l}) indicate well-aligned
$\gamma$-phase Na$_x$CoO$_2$ ({\it P63/mmc}). The minor peaks near
2-$\theta$ 18$^o$ and 34.5$^o$ are the (003) and (006) peaks of the
$\alpha$-NaCoO$_2$ ({\it R3}) impurity phase. The coexistence of the
$\gamma$ and $\alpha$ phases is intrinsic in the Na nonstoichiometry
region 0.85 $\leq x \leq$ 0.9.\cite{Lee06}} \label{fig:fig1}
\end{figure}

The single crystalline compound of Na$_{0.85}$CoO$_2$ was grown by
the traveling solvent floating zone method using an optical image
furnace.\cite{Chen04} The crystal was cleaved into a lamellar
hexagonal shape. The chemical concentration of Na$_{0.85}$CoO$_2$
was carefully analyzed by energy dispersive X-ray spectroscopy (EDX)
and inductively coupled plasma-atomic emission spectroscopy
(ICP-AES). The chemical distribution images by EDX confirmed that
the chemical inhomogeneity of the Na ion was less than 2 at.\%. The
single crystalline property was determined by using the Laue
back-scattering and X-ray diffraction techniques. Figure
\ref{fig:fig1} shows well-aligned peaks on the $c$-axis,
perpendicular to the crystal plane; the major sharp (00{\it l})
peaks indicate the $\gamma$-phase Na$_x$CoO$_2$ ({\it P63/mmc}). The
$c$-axis lattice parameter was determined to be $c=$10.71 {\AA} from
the powder X-ray diffraction measurements. We found a minor impurity
phase of $\alpha-$NaCoO$_2$ (\textit{R-3}) of less than 2 at.\%. The
double phases of $\gamma-$ and $\alpha-$Na$_x$CoO$_2$ were not
extrinsic impurity but intrinsic property in the Na range of 0.85
$\leq x \leq$ 0.95.\cite{Lee06} The minor phase (less than 2 at.\%)
of $\alpha-$NaCoO$_2$ might not affect the magnetic property because
NaCoO$_2$ is a nonmagnetic material. The temperature-dependent DC
magnetization $M(T,H)$ was measured by the magnetic property
measurement system (Quantum Design, U.S.A.) in a temperature range
of 2 K $\leq T \leq$ 300 K under fixed magnetic fields of $H=$ 0.1,
1.0, 3.0, and 5.0 T. The frequency- and temperature-dependent AC
magnetic susceptibilities $\chi(f,T)\equiv dM/dH$ were measured by
the Physical Property Measurement System (Quantum Design, U.S.A.)
with the AC magnetic susceptometer (ACMS) probe at low temperatures
($T\leq$ 20 K) and various frequencies of 0 Hz $\leq f \leq$ 10 kHz
with the excitation magnetic field $H_{ex}=$ 10 Oe.

\section{Result and Discussion}
\subsection{Static magnetic susceptibility}
\begin{figure}
\includegraphics [width=0.5\textwidth]{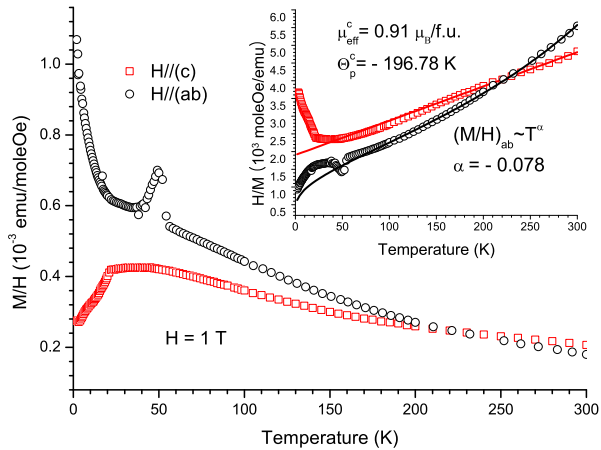}
\caption{(Color online) Temperature-dependent DC magnetic
susceptibility $(M/H)(T)$ of the Na$_{0.85}$CoO$_2$ single crystal
under a static magnetic field $H=$ 1 T for different crystal
orientations of $H\parallel$ (ab) (black circle) and $H\parallel$
(c) (red rectangle) in the temperature range of 2 K $\leq T \leq$
300 K. Inset shows temperature-dependent inverse DC magnetic
susceptibility $(H/M)(T)$. Lines are fitted results of Curie-Weiss
law at high temperatures 150 K $\leq T \leq$ 300 K for $H\parallel$
(c) (red line) and power-law behavior with $\chi\sim T^{\alpha}$ at
$T\geq$ 100 K for $H\parallel$ (ab) (black line). } \label{fig:fig2}
\end{figure}

Figure \ref{fig:fig2} presents the temperature-dependent static
magnetic susceptibility $M/H(T)$ of Na$_{0.85}$CoO$_2$ under a
magnetic field of $H=$ 1 T for different crystal orientations of
$H\parallel$ (ab) and $H\parallel$ (c) in the temperature range of 2
K $\leq T \leq$ 300 K. The magnetic susceptibility for $H\parallel$
(c)-axis $(M/H)_c(T)$ exhibits an antiferromagnetic (AF) transition
at $T_N\simeq$ 22 K, which is comparable with that in the case of
Na$_{0.83}$CoO$_2$ ($T_N\approx$ 20 K).\cite{Lin07} It follows the
Curie-Weiss law from 300 K to 150 K as shown in the inset of Fig.
\ref{fig:fig2} (red square, red line). From the Curie-Weiss fitting,
the effective magnetic moment and Weiss temperature are estimated to
be $\mu_{eff}=$ 0.91 $\mu_B$/f.u. and $\Theta_p=-$196.8 K,
respectively. Under the assumption of a low spin state (LS), the
magnetic moments of Co$^{4+}$ (3{\it d}$^5$, $S=$ 1/2) and Co$^{3+}$
(3{\it d}$^6$, $S=$ 0) are 1.732 $\mu_B$ and zero, respectively. If
we consider only the Co low spin-state, the partial fraction of
Co$^{4+}$ is 52 at.\%, which is identical to the average valence of
Co (3.52). The average Co valence, estimated by the effective
magnetic moment, is a little higher than both the $4-x$ value (3.15)
and the value determined by a redox titration.\cite{Maarit05} From
the wet chemical analysis of the Co valence for Na$_{0.85}$CoO$_2$,
the Co valence was estimated to be 3.15. The Co valence higher than
our estimation implies that the spin state of Co is not a low-spin
state. The spin state transition of the Co ion is widely observed in
the cobaltates, for example, perovskite cobaltates {\it
R}$_{1-x}${\it A}$_x$CoO$_3$ ({\it R}: rare earth and {\it A}:
alkaline earth elements).\cite{Wang06} In this system, the Co$^{3+}$
spin state is changed by the spin state transition from a low-spin
(LS, $t_{2g}^6$, $S=$ 0) to an intermediate-spin (IS,
$t_{2g}^5e_g^1$, $S=$ 1) or high spin state (HS, $t_{2g}^4e_g^2$,
$S=$ 2) with respect to temperatures. From the optical conductivity
measurement of the Na$_{0.82}$CoO$_2$ single crystal, the charge
ordering of Co$^{4+}$ and Co$^{3+}$ was revealed in the CoO$_2$
layers.\cite{Bernhard04} The mixed valence of Co$^{4+}$ and
Co$^{3+}$ in cobaltates easily invoke the spin state transition. The
Co$^{4+}$ ion lowers the local symmetry of the Co$^{3+}$ ion, which
induces the band splitting of the $e_g$ doublet and $t_{2g}$ triplet
states. If the energy gap between $e_g$ and $t_{2g}$ orbitals is
smaller than the Hund coupling, the intermediate- or high-spin
states are favorable because electrons can be transferred from the
highest $t_{2g}$ level to the lowest $e_g$ level. The intermediate-
or high-spin states of Co$^{3+}$ ions are not experimentally
observed directly. A direct experimental observation of the spin
state of Co in $\gamma$-Na$_x$CoO$_2$ system would be of
considerable interest.

\begin{figure}
\includegraphics [width=0.5\textwidth]{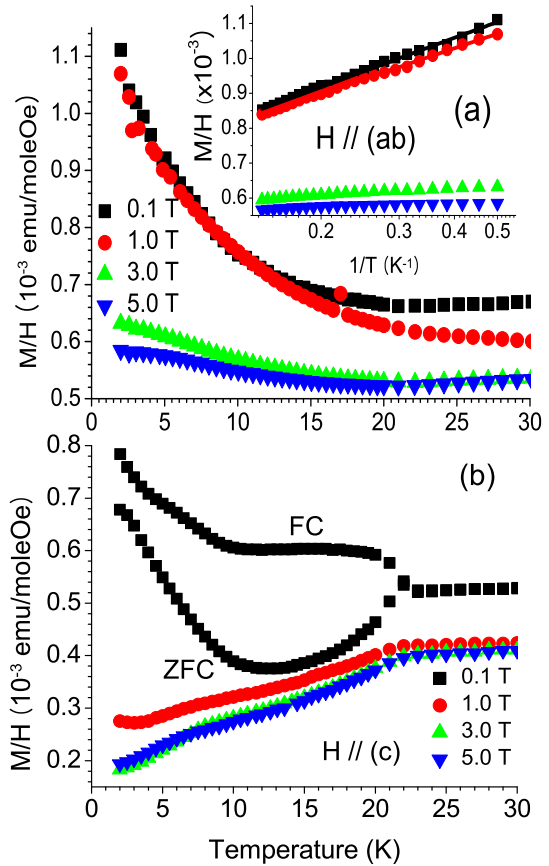}
\caption{(Color online) Temperature-dependent DC magnetic
susceptibility $(M/H)(T)$ at low temperatures (2 K $\leq T \leq$ 30
K) under static magnetic fields, as indicated, of $H\parallel$
(ab)-axis (a) and of $H\parallel$ (c)-axis (b), respectively. Inset
of (a) depicts the logarithmic divergent behavior of the DC magnetic
susceptibility $(M/H)_{ab}$ at low temperatures (2 K $\leq T \leq$
6.8 K) for various magnetic fields, depicted by the semilog plot of
$M/H$ vs. $1/T$.} \label{fig:fig3}
\end{figure}

The static magnetic susceptibility $(M/H)_{ab}(T)$ under a magnetic
field of $H=$ 1 T parallel to the (ab)-plane monotonically increases
with decreasing temperatures from 300 K to 60 K. With decreasing
temperatures, a peak is observed near $T\approx$ 50 K with a
subsequent increase in susceptibility at low temperatures ($T\leq$
10 K). Even though this peak (near 50 K) is sharper than that in the
other case,\cite{Lin07} the broad shoulder of $(M/H)_{ab}$ in the
intermediate temperature region (20 K $\leq T \leq$ 50 K) is
commonly observed for high Na concentration compounds of
Na$_{0.82}$CoO$_2$.\cite{Bayrakci04} The authors argue that this may
be a signal of a valence fluctuation between Co$^{3+}$ and
Co$^{4+}$.\cite{Rhyee} The inverse magnetic susceptibility
$(H/M)_{ab}(T)$ (black circle, black line), as shown in the inset of
Fig. \ref{fig:fig2}, follows the power-law behavior in the
temperature range of 100 K $\leq T \leq$ 300 K according to the
following equation: $\chi=\chi_0+AT^{\alpha}$, where the
zero-temperature susceptibility $\chi_0=$ 0.0028 emu/moleOe,
pre-factor $A=$ 0.0046 emu/moleOeK, and power exponent $\alpha=-$
0.078. The non-Curie-Weiss behavior of a magnetic ordered system is
an indication of the strongly correlated electron behavior. One
possible scenario of the power-law behavior in the magnetic
susceptibility is the multichannel Kondo effect.\cite{Stewart01} The
multichannel Kondo model is based on the strong correlation between
the local impurity spin and the conduction electron spin. The spins
of the conduction electron near the local impurities are paired with
a singlet state due to the antiferromagnetic coupling by Pauli
matrices.\cite{Schlottmann95} When the total number of electrons
overcompensates the local spin $n>2S$ (where $n$ is the number of
degenerate orbital channels and $S$ is the local spin), the
power-law or logarithmic divergent behaviors of magnetization,
resistivity, or specific heat are observed. The exact solution of
the multichannel Kondo model for dilute impurities show that the
magnetic susceptibility should have the power-law dependence with
temperatures: $\chi\propto T^{\alpha}$, where $\alpha=4/(n+2)-1$.
Under this consideration, the exponent parameter is estimated to be
$\alpha=-$0.078 from the magnetic susceptibility $(M/H)_{ab}(T)$ at
high temperatures ($T\geq$ 100 K). The total number of electron
channels $n$ is obtained as $n=$ 2.34, which is the case of
overcompensation $n>2S$ (where $S=$ 1/2 or 1). The multichannel
Kondo effect can be manifested at high temperatures ($T\geq
T_K$).\cite{Pantke98} In order to clarify the unconventional
power-law behavior of $(M/H)_{ab}(T)$ at high temperatures ($T\geq$
100 K), more detailed investigations need to be carried out by using
the electrical resistivity, heat capacity, and thermopower
measurements.

The logarithmic divergent behavior of $(M/H)_{ab}(T)$ is observed at
low temperatures ($T\leq$ 7 K), which is different from the Fermi
liquid behavior. It may be argued that the impurity spin effect of
$(M/H)_{ab}(T)$ is significant due to the symmetrically coupled
half-integer spin impurity with the valence fluctuation. Figure
\ref{fig:fig3} (a) shows the temperature-dependent magnetic
susceptibility $(M/H)_{ab}(T)$ for different magnetic fields as
indicated with $H\parallel$ (ab). The $(M/H)_{ab}(T)$ increases
logarithmically at low temperatures (2 K $\leq T\leq$ 7.5 K) as
shown in the inset of Fig. \ref{fig:fig3}(a). When an impurity spin
$S$ is coupled with an antiferromagnetic background, the sublattice
symmetric impurity spin $S=$ 1/2 forms a collective spin state of
spinon by pulling two spins, resulting in an orthogonal pair of
degenerated ground states. If the symmetric impurity spin 1/2, which
is coupled with an antiferromagnetic ground state, is dressed by the
valence fluctuation, the impurity spin susceptibility would have a
non-Curie divergence given by the following relation:
$\chi=\chi_0+\ln(T^*/T)/T^*$, where $T^*$ ($T\ll T^*$) is the
characteristic temperature with the energy scale of $T^*\propto
J\exp(-ConstJ/g)$ for weak coupling ($g\ll J$) ($J$ is the transfer
integral and $g$ is the coupling strength).\cite{Clarke93} From the
fitting of the above relation, the zero temperature susceptibility
$\chi_0$ and characteristic temperature $T^*$ are estimated to be
$\chi_0=$ 0.0003 emu/moleOe and $T^*=$ 10,451.02 K $=$ 0.90 eV for
$H=$ 0.1 T and $\chi_0=$ 0.0004 emu/moleOe and $T^*=$ 5,263.16 K $=$
0.45 eV for $H=$ 1.0 T. For high magnetic fields ($H=$ 3 T and 5 T),
the above relation is not fitted. The spin impurity effect is
suppressed with increasing magnetic fields as shown in Fig.
\ref{fig:fig3}(a). The suppression of the impurity spin effect with
increasing magnetic fields might have been caused by the decreasing
coupling strength $g$ between the impurity spin and the
antiferromagnetic background.

Figure \ref{fig:fig3}(b) depicts the magnetic susceptibility
$(M/H)_c$ for $H\parallel$ (c)-axis at low temperatures (2 K $\leq
T\leq$ 30 K) for different magnetic fields as indicated. The
antiferromagnetic transition $T_N=$ 22 K is not changed
significantly with increasing magnetic fields. The important feature
is the strong thermal hysteresis between the zero-field-cooled (ZFC)
and the field-cooled (FC) magnetic susceptibility at $H=$ 0.1 T as
shown in the Fig. \ref{fig:fig3}(b). While the N\'{e}el temperature
$T_N=$ 22 K is relatively low, the derived value of the Weiss
temperature $\Theta_p=-$ 196.78 K from the Curie-Weiss fitting is
very large, which implied a strong antiferromagnetic interaction.
The significant discrepancy between the N\'{e}el temperature and the
Weiss temperature is the evidence of the spin frustration. The
frustration parameter, defined by $f=|\Theta_p /T_N|$, is equal to
8.9, which is much larger than that of a conventional
antiferromagnet. The antiferromagnetic compounds of Na$_x$CoO$_2$
may have the geometrical spin frustration for a wide range of Na
concentrations ($x\geq$ 0.78) due to the triangular crystal
structure of the CoO$_2$ sublattice. The $\mu$SR study of
Na$_x$CoO$_2$ ($x=$ 0.78, 0.87, and 0.92) provided the evidence for
the strong spin frustration.\cite{Bernhard07} A small thermal
hysteresis between ZFC and FC is also observed for the magnetic
susceptibility $(M/H)_{ab}(T)$ with applied magnetic field in the
$H\parallel$ (ab)-plane at low temperatures ($T\leq$ 20 K) under a
magnetic field of $H=$ 0.1 T (not shown here). The thermal
hysteresis between ZFC and FC for different crystal orientations
implies the unstable antiferromagnetic background of the
Na$_{0.85}$CoO$_2$ compound.

\begin{figure}
\includegraphics [width=0.5\textwidth]{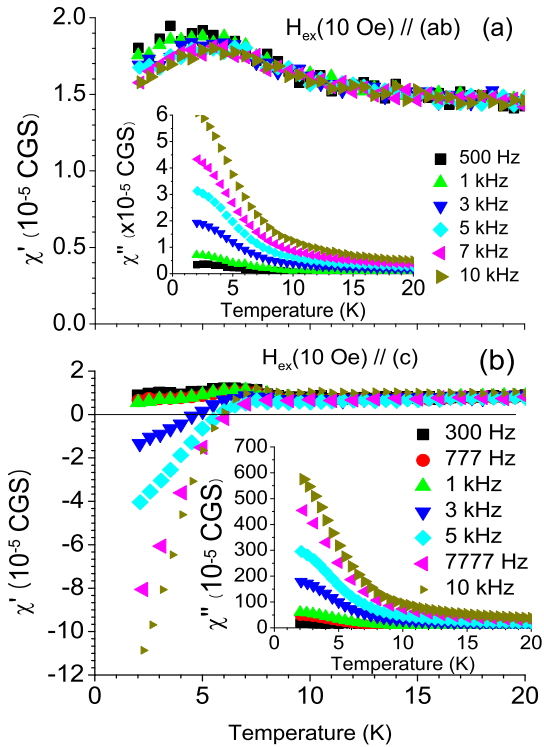}
\caption{(Color online) Temperature-dependent real part of the AC
magnetic susceptibility $\chi'(T)$ for different excitation field
($H_{ex}=$ 10 Oe) frequencies $f$, as indicated, for
$H_{ex}\parallel$ (ab)-axis (a) and $H_{ex}\parallel$ (c)-axis (b),
respectively, at low temperatures (2 K $\leq T \leq$ 20 K). Insets
of (a) and (b) are the imaginary part of the AC magnetic
susceptibility $\chi''(T)$ for the corresponding excitation field
frequencies and crystal orientations.} \label{fig:fig4}
\end{figure}

\subsection{Dynamic magnetic susceptibility}
In order to study the dynamic spin behavior of this compound, we
measured the AC magnetic susceptibility $\chi\equiv
dM/dH=\chi'+i\chi''$ for the crystal orientations of both
$H\parallel$(ab)- and $H\parallel$(c)-axis. Figure \ref{fig:fig4}(a)
and (b) represent the temperature-dependent real $\chi'$ and
imaginary $\chi''$ (insets) parts of the AC magnetic susceptibility
for various magnetic field frequencies as indicated with different
crystal orientations of $H_{ex}\parallel$ (ab)- and (c)-directions,
respectively, where $H_{ex}$ is the excitation AC magnetic field of
10 Oe. Here, the demagnetization factor ($N=$ 1) for $H\parallel$(c)
was considered for the following equations because we used the
plate-like single crystal.
\begin{equation}
\chi_m = {dM \over dH_a} = {\chi \over 1+N\chi} = \chi'_m +
i\chi''_m
\end{equation}

\begin{equation}
\chi = {dM \over dH} = {\chi_m \over 1-N\chi_m} = \chi' + i\chi''
\end{equation}

\begin{equation}
\chi' = {\chi'_m -N({\chi'_m}^2+{\chi''_m}^2) \over
N^2({\chi'_m}^2+{\chi''_m}^2)-2N\chi'_m+1} \\
\end{equation}

\begin{equation}
\chi'' = {\chi''_m \over N^2({\chi'_m}^2+{\chi''_m}^2)-2N\chi'_m+1}
\end{equation}

\noindent where $\chi_m$ and $\chi$ are the measured and intrinsic
AC magnetic susceptibilities, respectively.

In the plot of the dynamic magnetic susceptibility $\chi_{ab}$ for
the $H\parallel$(ab)-plane, as shown in Fig. \ref{fig:fig4}(a), a
broad peak is observed near $T_g\approx$ 4.5 K for the magnetic
field frequency $f=$ 500 Hz. With increasing magnetic field
frequency, the temperatures at which the broad peak is observed
increase slightly from 4.5 K for $f=$ 500 Hz to 6.0 K for $f=$ 10
kHz. The frequency-dependent antiferromagnetic-like transition of
the real part of the AC magnetic susceptibility $\chi'(f,T)$ is not
a characteristic of an antiferromagnet but a characteristic of the
spin glass-like behavior. The AC magnetic susceptibility behavior
$\chi'_{ab}$ of the glass-like transition at $T_g\approx$ 4.5 K
without a phase transition near the antiferromagnetic temperature
$T_N=$ 22 K is similar to that observed in the case of the
Na$_{0.71}$CoO$_2$ compound.\cite{Wooldridge05} The glass freezing
temperatures are defined by the cusp points of $\chi'_{ab}$. It is
not a conventional spin glass system because of the strong energy
dissipation of the AC magnetic susceptibility. The imaginary part of
the AC magnetic susceptibility ($\chi''_{ab}$) along the (ab)-plane
increases with decreasing temperatures and increasing magnetic field
frequencies as shown in the inset of Fig. \ref{fig:fig4}(a). The
significant increase in $\chi''_{ab}$ with respect to decreasing
temperatures and increasing frequencies is an uncommon behavior of
the spin glass system. The out-of-phase part $\chi''$ is much larger
than the in-phase part $\chi'$ at low temperatures. In a
conventional spin glass system, the $\chi''$ has broad peaks near
the spin glass transition with a weaker signal than that of
$\chi'$.\cite{Binder86,Hanasaki07} However, in this compound, the
energy dissipation part ($\chi''$) of the dynamic spin
susceptibility significantly enhanced with decreasing temperatures
and increasing frequencies. Therefore, it is concluded that this
system is not a simple spin glass system but an exotic system that
exhibits strong energy dissipation. Moreover, the real part of the
AC magnetic susceptibility $\chi'_c$ along the (c)-direction
$H\parallel$(c) exhibits an unconventional sign change from positive
to negative with decreasing temperatures ($T\leq$ 7 K) and
increasing excitation field frequencies ($f\geq$ 3 kHz) as shown in
Fig. \ref{fig:fig4}(b). This system is quite different from that of
the Na$_{0.71}$CoO$_2$ compound as well as the conventional spin
glass system.\cite{Wooldridge05} In an usual case, the negative sign
of $\chi'$ implies a diamagmetic signal. However, this compound is
neither a diamagnet nor a superconductor. First, it displays the
strong antiferromagnetic transition at $T_N=$ 22 K for the
$H\parallel$(c)-axis as shown in Fig. \ref{fig:fig3}(b). Second,
like $\chi''_{ab}$, the energy dissipation of magnetic
susceptibility $\chi''_c$ is very significantly large. Compared to
$\chi''_{ab}$ (6.0$\times$10$^{-5}$ dimensionless in CGS unit at 2 K
and 10 kHz), the value of $\chi''_c$ (5.8$\times$10$^{-3}$) is
larger by two orders of magnitude than that of $\chi''_{ab}$. In
superconducting materials, the energy dissipation of the dynamic
spin susceptibility $\chi''$ diminishes below the transition
temperature.

\begin{figure}
\includegraphics [width=0.5\textwidth]{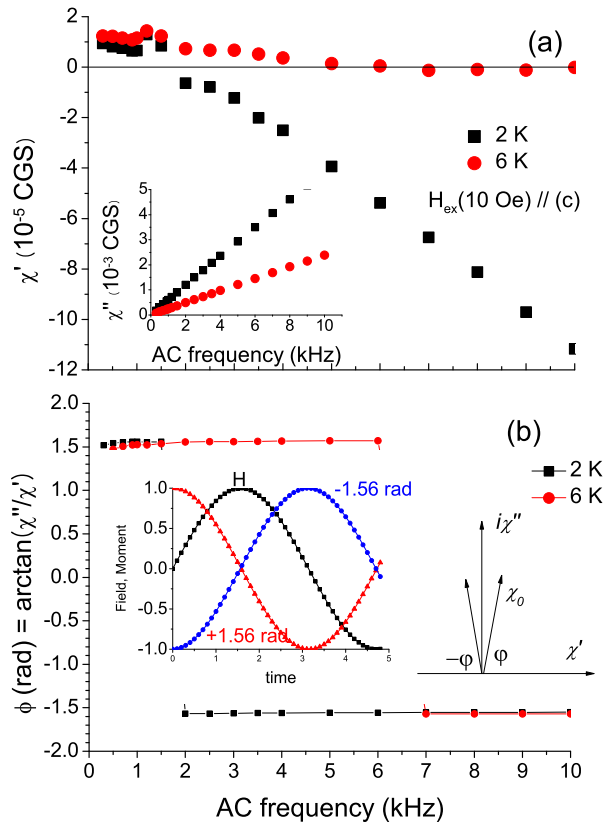}
\caption{(Color online) (a): Isothermal frequency-dependent real
part $\chi'(f)$ and imaginary part $\chi''(f)$ (inset) AC
susceptibilities at constant temperatures $T=$ 2 K and 6 K under the
excitation field $H_{ex}=$ 10 Oe parallel to (c)-direction. (b):
Isothermal frequency-dependent phase difference $\phi(f)$, defined
by $\arctan(\chi''/\chi')$, at $T=$ 2 K and 6 K. Insets of (b)
depicts the concept of the sign change of the phase difference
$\phi$ of the dynamic magnetic susceptibility. Right inset of (b) is
the magnitude $\chi_0$ and the phase difference $\phi$ of the
dynamic susceptibility in the complex phase space. Left inset of (b)
represents the applied magnetic field (black rectangle) and derived
magnetic susceptibility of in-phase (red triangle) and out-of-phase
(blue circle) response with time.} \label{fig:fig5}
\end{figure}

In order to understand the unusual dynamical spin susceptibility
behavior of $\chi_c$, we measured the isothermal frequency-dependent
AC susceptibility $\chi_c(f)$ at the temperatures of 2 K and 6 K, as
shown in Fig. \ref{fig:fig5}(a). With increasing AC field frequency
$f$ at $T=$ 2 K, $\chi'_c(f)$ decreases and the sign changes at
$f\approx$ 1.8 kHz. The small peaks at $f=$ 1.2 kHz for $T=$ 2 K and
6 K have not been understood as yet. The imaginary part of the AC
susceptibility $\chi''_c(f)$ increases linearly with increasing AC
field frequency, as shown in the inset of Fig. \ref{fig:fig5}(a),
which means the linear increase in energy dissipation. The
out-of-phase part of the AC susceptibility $\chi''_c$ is much more
significant than the in-phase part of $\chi'_c$. The sign change of
$\chi'_c(f)$ from positive to negative with increasing AC field
frequency can be understood by the abrupt sign change of the phase
difference $\phi$ with respect to the AC field frequency. From the
definition of AC susceptibility $\chi\equiv \chi'+i\chi'' = \chi_0
\cos\phi + i\chi_0\sin\phi$, the magnitude of AC susceptibility
$\chi_0$ and the phase difference $\phi$ can be represented by
$\chi_0=\sqrt{\chi'^2+\chi''^2}$ and $\phi=\arctan(\chi''/\chi')$,
respectively. The phase difference $\phi$ is the relative phase
shift between the applied AC magnetic field and the derived magnetic
signal. Because the $\chi''$ increases linearly with increasing
field frequency $f$ and the orders of magnitude of $\chi''$ is much
larger than that of $\chi'$, $\chi_0$ increases linearly with AC
field frequencies (not shown). The phase difference $\phi$ exhibits
an abnormal abrupt sign change from 1.56 rad (89.4$^o$) to $-$1.56
rad (90.6$^o$) at the frequencies of the $\chi'_c$ sign change, as
shown in Fig. \ref{fig:fig5}(b). The schematic view of the magnetic
susceptibility in the complex phase space is shown in the inset of
Fig. \ref{fig:fig5}(b). At low frequencies ($f\leq$ 1.8 kHz at $T=$
2 K), these insets exhibit the in-phase response of magnetic
susceptibility (red triangle) to the applied AC magnetic field
(black rectangle) with $\phi=$ 1.56 rad. However, at high
frequencies, the phase difference $\phi$ abruptly changes to $-$1.56
rad (blue circle) due to a significant increase in the out-of-phase
response $\chi''$. The in-phase to out-of-phase shift of the
response of the AC magnetic susceptibility may be a result of the
strong energy dissipation $\chi''_c$ of this compound. Due to the
fact that $\chi''_c$ is much larger than $\chi'_c$, the phase
difference $\phi$ is on the boundary between the in-phase and the
out-of-phase regions in the magnetic susceptibility complex space
($\phi\approx\pi/2$ rad) (right inset of Fig. \ref{fig:fig5}(b)). At
present, why the imaginary part of the magnetic susceptibility
$\chi''_c$ is much stronger than the real part of the susceptibility
$\chi'_c$ is not clear. The dynamical spin property of this compound
may be related to the strong spin frustration in the
$H\parallel$(c)-axis.

\section{Conclusion}
In summary, we synthesized a high-quality single crystalline
compound of Na$_{0.85}$CoO$_2$ by the travelling solvent floating
zone method. It was observed from the in-plane and the out-of-plane
measurements that the DC magnetic susceptibility showed significant
crystal anisotropic properties upon the application of a magnetic
field. The DC magnetic susceptibility along the c-axis followed the
typical Curie-Weiss behavior with a stable antiferromagnetic
transition at $T_N=$ 22 K. The strong thermal instability between
the zero-field-cooled (ZFC) and the field-cooled (FC) magnetization
and the high frustration index indicated a strong spin frustration
along the c-axis. The in-plane magnetization exhibited the anomalous
power-law behavior at high temperatures ($T\geq$ 100 K) and the
logarithmic divergent behavior ($\propto\ln(T^*/T)$) at low
temperatures ($T\leq$ 10 K). The high temperature power-law behavior
may be related with the multichannel Kondo effect. The logarithmic
divergence can be understood by the impurity spin effect, dressed by
the valence fluctuation. Remarkable behavior of the AC magnetic
susceptibility was observed at high excitation field frequencies
($f\geq$ 3 kHz) along the c-axis. The sign of the real part of the
AC magnetic susceptibility along the c-axis changed from positive to
negative with decreasing temperatures (near 7 K). We argued that the
sign change may due to the abrupt sign change of the phase
difference between the applied AC magnetic field and the magnetic
susceptibility from 1.56 rad (89.4$^o$) to $-$1.56 rad (90.6$^o$).
Because the imaginary part of the AC magnetic susceptibility was
significantly larger than the real part one, the phase difference
was on the boundary between the in-phase and the out-of-phase
regions. A fundamental question remains as to why the energy
dissipation was so significant at high frequencies and low
temperatures of this compound. Further investigations would help in
clarifying this behavior.

\acknowledgments We would like to express our sincere gratitude to
G. G\"{o}tz and Christof Busch for their technical assistance.
{}


\end{document}